\documentclass{PoS}

\title{Searches for Point Sources of High Energy Cosmic Neutrino with the ANTARES Telescope}

\ShortTitle{Neutrino Point Source Searches}

\author{\speaker{Dornic D.} on behalf of the ANTARES Collaboration\\
        CPPM, Aix-Marseille Universit\'ee, CNRS/IN2P3, Marseille, France\\
	IFIC - Instituto de Fisica Corpuscular, Edificios Investigaci\'on de Paterna, CSIC - Universitat de Valencia, Apdo. de Correos 22085, 46071 Valencia, Spain\\
        E-mail: \email{dornic@cppm.in2p3.fr}}

\abstract{The ANTARES observatory is currently the largest neutrino telescope in the Northern 
Hemisphere. It is well suited to detect high energy neutrinos produced in astrophysical sources 
as it can observe a full hemisphere of the sky at all the times with a high duty cycle and 
an angular resolution about 0.4 degrees. Due to its location in the South of France, 
ANTARES is sensitive to up-going neutrinos from many potential galactic sources in the TeV to 
PeV energy regime. Results from a time-integrated unbinned method as well as the sensitivity of 
the detector using 2007-2010 data are presented. Moreover, using a time-dependent search for the 
transient sources, the background rejection and point-source sensitivity can be drastically 
improved by selecting a narrow time window around the assumed neutrino production period. The 
gamma-ray light curves of blazars measured by the LAT instrument on-board the Fermi satellite 
reveal important time variability information. A strong correlation between the gamma-ray and 
the neutrino fluxes is expected in a hadronic scenario. First results on the search for ten 
bright and variable Fermi sources with the 2008 ANTARES data are also presented.}

\FullConference{The 2011 Europhysics Conference on High Energy Physics-HEP 2011,\\
		July 21-27, 2011\\
		Grenoble, Rh\^one-Alpes France}

\begin{document}

\section{Introduction}
Neutrinos are unique messengers to study the high-energy universe as they are neutral and 
stable, interact weakly and therefore travel directly from their point of creation to the 
Earth without absorption. Neutrinos could play an important role in understanding the 
mechanisms of cosmic ray acceleration and their detection from a cosmic source would be a 
direct evidence of the presence of hadronic acceleration. The production of high-energy 
neutrinos has been proposed for several kinds of astrophysical sources, such as active 
galactic nuclei (AGN), gamma-ray bursters (GRB), supernova remnants and microquasars, in 
which the acceleration of hadrons may occur (see Ref.~\cite{bib:Becker} for a review). 

The ANTARES Collaboration has constructed a neutrino telescope~\cite{bib:Antares} at a depth 
of about 2475 meters, offshore Toulon, France. Neutrinos are detected by photomultiplier 
tubes (PMTs), housed in pressure resistant glass spheres, which are regularly 
arranged on 12 detection lines. Each line accommodates up to 25 triplets of PMTs, located 
between 100 and 450 m above the sea bed. The lines are connected to the shore via a junction 
box and a single, 40 km electro-optical cable, which provides both power and an optical data 
link. On shore, a computer farm runs a set of trigger algorithms to identify events containing 
Cherenkov light from high energy muons within the data stream, which otherwise consists mostly 
of signals from radioactive decay and bioluminescence. In 2007, the first 5 detector lines became 
operational, followed, in May 2008, by the completion of the full 12-line detector.

Most of the analyses described here use a muon track reconstruction algorithm (based on 
Ref.~\cite{bib:AAfit}) that consists of multiple fitting steps. The final step is based on a 
full likelihood description of the arrival times of the detected Cherenkov photons, which also 
accounts for background light. The achieved angular resolution is, by necessity, determined 
from simulations. However, several aspects of the simulations were confronted with data in 
order to constrain the possible systematic effects in the timing resolution that would result 
in a deteriorated angular resolution. The angular resolution (i.e. the median angle between the 
neutrino and the reconstructed muon) was found to be 0.4$\pm$0.1 (sys) degrees for the detector 
with all 12 lines operational. 

\section{Time-Integrated Point Source Search}
Cosmic point-like source of neutrinos have been searched for using 813 live-days of data from 
2007 up to and including 2010. An earlier version of the analysis is described in 
Ref.~\cite{bib:AAfitps}. Event selection criteria have been applied which optimize both the 
sensitivity and the discovery potential. Events are required to be reconstructed as upward going 
and to have a good reconstruction quality, quantified by a variable based on the reduced log-likelihood 
of the track fit, and an angular error estimate better than 1$^{\circ}$. The resulting event sample 
consists of 3058 neutrino candidates, of which about 84(16)\% is expected to be amospheric neutrinos 
(muons misreconstructed as upward-going). To search for point sources, the analysis uses an unbinned 
maximum likelihood method, which exploits the knowledge on the very good angular resolution 
and the rate of background events as a function of the declination. 

\begin{figure}[ht!]
\centering
\includegraphics[width=0.45\textwidth]{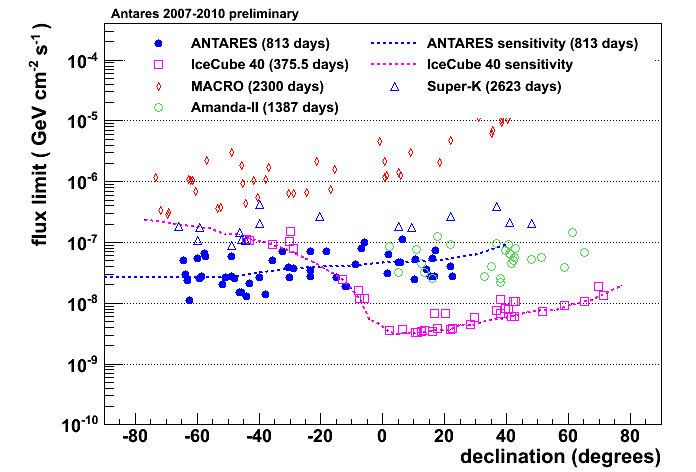}
\caption{Limits set on the normalisation of an E$_{\nu}^{-2}$ spectrum of high energy neutrinos from 
selected candidates (blue plein circles). Also shown is the sensitivity (dashed blue line), which is defined as the median expected limit. In 
addition to the present result, several previously published limits on sources in both the Southern 
and Northern sky are also shown.}
\label{fig:limits}
\end{figure}

Two different versions of the search 
were conducted: in the full-sky search, the full visible sky is searched for point sources. In the 
candidate search, neutrinos are searched for only in the direction of 24 a-priori selected candidate 
source locations, corresponding to known gamma ray objects of interest. Neither search yields a 
significant excess of events over the background: the post-trial p-values are 2.5\% (for a cluster 
of events at $\alpha,\delta$ = (-46.5$^{\circ}$ , -65.0$^{\circ}$) for the full sky search and 41\% 
for the most signal-like source in the candidate source list (HESS J1023-575). Limits have been 
extracted on the intensity of an assumed E$^{-2}$ neutrino flux from the candidate sources. They 
are shown in Figure~\ref{fig:limits}. The limit computation is based a large number of 
generated pseudo experiments in which systematic uncertainties on the angular resolution and 
acceptance are taken into account. These limits are more stringent than those from previous
experiments in the Northern hemisphere (also indicated in the figure) and competitive with those 
set by the IceCube Observatory~\cite{bib:IceCube} for declinations < -30$^{\circ}$. The various
experiments are sensitive in different energy ranges, even though they all set limits on E$^{-2}$ 
spectra. For this spectrum, ANTARES detects most events at energies in a broad range around 10 TeV, 
which is a relevant energy range for several galactic source candidates.

\section{Time-Dependent Point Source Search}
In addition to the time-integrated searches described above, a time-dependent point source search has 
been conducted to look for neutrinos in correlation to the variable gamma-ray emission from blazars 
measured by the LAT instrument on board the Fermi satellite. By restricting the search to the high state
 (typically 1-20 days) of the gamma emission, the background is reduced and thus, the sensitivity to 
 these sources has been improved by about a factor about two with respect to a standard time-integrated 
 point source search.  

The ANTARES data used in this analysis were taken in the period from September 6 to December 31, 2008 
(54720 to 54831 modified Julian days, MJD) with the twelve line detector. This period overlaps with the 
availability of the first data from the LAT instrument onboard the Fermi satellite. The resulting sample 
consists of 628 events obtained in the corresponding effective live time of 60.8 days.

\begin{figure}[ht!]
\centering
\includegraphics[width=0.45\textwidth]{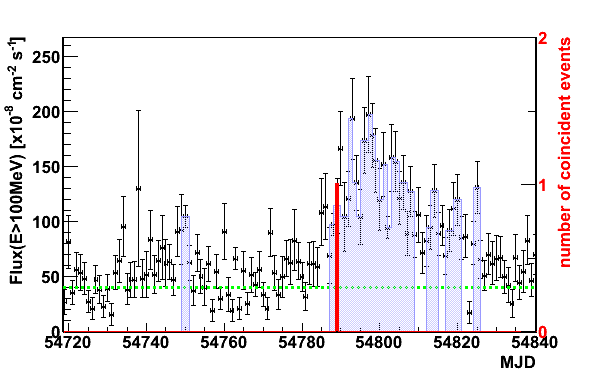}
\caption{Gamma-ray light curve (dots) of the blazar 3C279 measured by the LAT instrument onboard the Fermi 
satellite above 100 MeV. The light shaded histogram (blue) indicated the high state periods. The dashed line 
(green) corresponds to the fitted baseline. The red histogram displays the time of the associated ANTARES 
neutrino event.}
\label{fig:Result_3C279}
\end{figure}

This search was applied to ten very bright and variable Fermi LAT blazars. For nine sources, no coincidences 
are found. For 3C279, a single high-energy neutrino event is found in 
coincidence. This event is located at 0.56$^{\circ}$ from the source location during a large flare in 
November 2008. The pre-trial p-value is 1.0~\%. Figure~\ref{fig:Result_3C279} shows the time distribution 
of the Fermi gamma-ray light curve of 3C279 and the time of the coincident neutrino event. This event was 
reconstructed with 89 hits distributed on ten lines with a track fit quality $\Lambda=-4.4$. The post-trial 
probability computed taking into account the ten searches is 10~\% and is thus compatible with background 
fluctuations.

\section{Summary}

The first deep-sea neutrino telescope, ANTARES, has been taking data for four and a half years now. A large 
number of analysis are being performed, looking for astrophysical signals of neutrinos, either stand-alone 
or by looking for coincident observations with a variety of other experiments. The geographical position, 
combined with the good angular resolution allow ANTARES to explore, in particular, Galactic neutrino sources 
in the relevant energy range. Neither the time-integrated nor the time-dependent search show a significant excess of
events.

\section{Acknowledgments}
I greatfully acknowledge the financial support of MICINN (FPA2009-13983-C02-01 and MultiDark 
CSD2009-00064) and of Generalitat Valenciana (Prometeo/2009/026).

\end{document}